\documentclass[%
 reprint,
superscriptaddress,
 amsmath,amssymb,
 aps,
pra,
]{revtex4-2}

\usepackage{graphicx}
\usepackage{dcolumn}
\usepackage{bm}
\usepackage{color}
\usepackage[hidelinks]{hyperref}

\begin{document}

\preprint{APS/123-QED}

\title{Condensate-mediated shape transformations of cellular membranes by capillary forces}

\author{Lukas Hauer}
\email{lukas.hauer@uni-koeln.de}
\affiliation{University of Cologne, Faculty of Medicine and University Hospital Cologne, Cologne, Germany}
\affiliation{Humboldt University of Berlin, Institute of Biology, 10115 Berlin, Germany}

\author{Katharina Sporbeck}%
\affiliation{Humboldt University of Berlin, Institute of Biology, 10115 Berlin, Germany}

\author{Joseph F. McKenna}%
\affiliation{School of Life Sciences, University of Warwick, Coventry CV4 7AL, United Kingdom}

\author{Dmytro Puchkov}%
\affiliation{Leibniz-Forschungsinstitut für Molekulare Pharmakologie, Berlin, Germany}

\author{Alexander I. May}%
\affiliation{Institute for Integrated Research, Institute of Science Tokyo, Tokyo, Japan}

\author{Lorenzo Frigerio}%
\affiliation{School of Life Sciences, University of Warwick, Coventry CV4 7AL, United Kingdom}

\author{Roland L. Knorr}
\affiliation{University of Cologne, Faculty of Medicine and University Hospital Cologne, Cologne, Germany}
\affiliation{Humboldt University of Berlin, Institute of Biology, 10115 Berlin, Germany}
\affiliation{Graduate School and Faculty of Medicine, The University of Tokyo, Tokyo 113-0033, Japan}%

\author{Amir H. Bahrami}
\email{bahrami@unam.bilkent.edu.tr}
\affiliation{UNAM-National Nanotechnology Research Center and Institute of Materials Science \& Nanotechnology, Bilkent University, Ankara, Turkey}

\date{\today}

\begin{abstract}
Phase-separated biomolecular condensates with liquid-like properties play a key role in the organization and compartmentalization of the intracellular environment. Condensate-mediated capillary forces acting on membranes drive physiologically important reshaping of membrane-bound organelles, such as vacuoles and autophagosomes. Here, we explore condensate-mediated membrane shape transformations. We employ {\textit{in planta}} live-cell imaging, an \textit{in vitro} reconstitution system with tunable interfacial tension, and computer simulations of an elastic membrane model to describe three morphologies of membrane structures localized at condensate interfaces: tubes, sheets, and cups. We find that the forces associated with high interfacial tension drive the formation of stable sheets, while tubes and cups prevail at lower interfacial tension. We calculate the free energies of each membrane shape and identify the energy barriers that govern the transitions between the shapes. With this approach, we find that shape transformations depend on the history of the interfacial membrane and exhibit a tube-to-cup hysteresis. These findings indicate that temporal control of condensate surface properties can mediate the morphogenesis of cup-like structures in cells, such as the formation of "bulbs" within plant vacuoles. Our results further generalize how the interplay of condensates and membranes contributes to intracellular organization. 

\end{abstract}

                            
\maketitle

\section*{Significance}
Cells use membraneless compartments, called condensates, to organize biochemical processes. We show that condensates can physically reshape cellular membranes, producing tubes, sheets, or cup-like structures by tuning the shape metastability through the system’s free energy. Using live-cell imaging, reconstituted systems, and computer simulations, we explain how surface tension controls these transformations and identify energy barriers that dictate shape changes. Our findings reveal a general mechanism by which condensates remodel membranes, providing insight into how plant cells can generate cup-like structures called bulbs. This work highlights the physical principles that underlie dynamic cellular organization.

\section{Introduction}
Compartmentalization within eukaryotic cells is vital for life and is achieved by membrane-enclosed organelles and liquid-like condensates. Condensates form {\textit{via}} liquid-liquid phase separation (LLPS) of biomolecules, including proteins and RNA in the cytosol or within membrane-bounded organelles \cite{brangwynne_germline_2009,franzmann_phase_2018,dorone_prionprotein_2021,staples_phase_2023}. Wetting by condensates results in capillary forces that can drive remodeling of cellular structures such as the cytoskeleton and membrane-enclosed organelles \cite{kusumaatmaja_intracellular_2021,gouveia_capillary_2022,boddeker_nonspecific_2022,fang_wet_2025}; since organelle shape is linked to organelle function \cite{shibata_mechanisms_2009,mathur_review_2020,jenkins_mitochondria_2024},  wetting thereby has potentially important functional implications.  
Recent studies have demonstrated the consequences of condensate-mediated organelle remodeling on the physiology of the cell. Examples include cytosolic condensates that control autophagosome closure \cite{agudo-canalejo_wetting_2021}, vacuolar condensates that drive the vesiculation of plant vacuoles \cite{kusumaatmaja_wetting_2021}, condensate-mediated endocytosis \cite{bergeron-sandoval_endocytic_2021}, and multivesicular body biogenesis \cite{wang_biomolecular_2024}. 

The morphology of condensate-free organelles in equilibrium is well understood in terms of minimizing the elastic deformation energy of the membrane \cite{helfrich_elastic_1973}. Typically, deformations result in three characteristic membrane shapes: tubes (prolates), sheets (oblates), and cups (stomatocytes) \cite{seifert_configurations_1997}. For example, the endoplasmic reticulum can be understood as a network combining tubular and sheet-like regions \cite{shibata_mechanisms_2010}. The Golgi apparatus is made up of membrane sheet stacks that form from tube-like intermediates at the cis-Golgi \cite{tachikawa_golgi_2017}. Autophagosomes are cup-shaped structures that form by the bending of expanding sheet-like precursor membranes \cite{knorr_curvature_2012}. The equilibrium shape of the organelle is determined primarily by the volume-to-area ratio \cite{seifert_shape_1991,seifert_configurations_1997}: decreasing this ratio drives shape transitions from tubes \textit{via} sheets to cups. On the other hand, a variety of biomolecules modulate the mechanical properties of membranes and their propensity to deformation \cite{doole_cholesterol_2022}. For example, insertion of amphipathic helices or hairpins from proteins into one side of the bilayer generates asymmetry and thereby induces spontaneous curvature in the membrane \cite{mcmahon_membrane_2005}.

Recently, condensate wetting was identified as an independent mechanism that modulates organelle shapes in cells \cite{kusumaatmaja_intracellular_2021, gouveia_capillary_2022}. In general, the shape of wet cellular substrates is governed by the elastocapillary, which describes the coupling between the substrate elasticity and the capillary force. The latter arises from interfacial tension, which preserves the integrity and spherical geometry of the condensate. Furthermore, the wetting affinity between condensate and substrate, quantified by the wetting contact angle, determines how strongly the interfacial tension acts on the substrate \cite{mora_capillarity_2010, graham_liquidlike_2023}. Wetting condensates with high interfacial tension stabilize membrane sheets \cite{agudo-canalejo_wetting_2021, zhao_membrane_2024}, thus preventing autophagosome formation \cite{agudo-canalejo_wetting_2021}.

\begin{figure*}
    \centering
    \includegraphics[width=.9\textwidth]{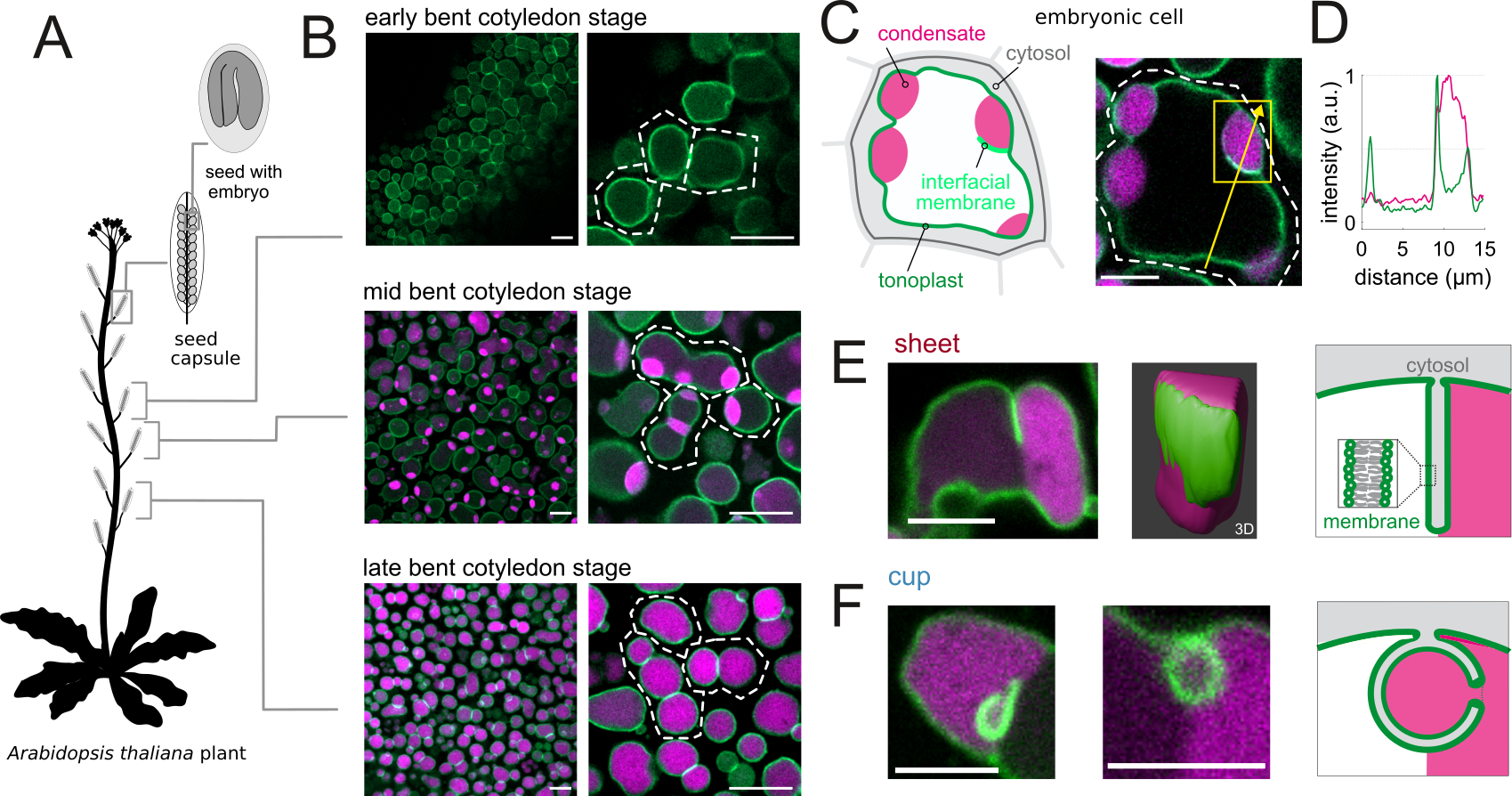}
    \caption{\label{fig:figure_1} Condensates within plant vacuoles shape internal membrane structures.
    (A) Each individual {\textit{A. thaliana}} plant can be used to sample multiple developmental stages of their embryos. Each fruit (seed capsule) contains 30-60 seeds (light gray) with a single embryo. (B) Embryo cells (white dashed lines to guide the eye, see \cite{feeney_protein_2018}) at different developmental stages are characterized by single, large vacuoles without condensates (early bent cotyledon stage), single large vacuoles with accessible condensate interfaces (mid bent cotyledon stage), and several small, condensate-filled vacuoles (late bent cotyledon stage). Confocal live-cell imaging of embryonic cotyledon (leaf) cells expressing the tonoplast protein marker TPK-GFP (membrane, green) and condensates (magenta, Neutral Red). (C) Single plant cell at mid bent cotyledon stage with vacuole (largest cellular organelle, green) and condensates (magenta). Left, schematic of plant cell with condensates and interfacial membrane. Right, corresponding confocal live-cell image. Yellow box, condensate with interfacial sheet. (D) Fluorescence intensity line profile along the yellow arrow in (C). (E) Interfacial sheets. Left, confocal section. Centre, 3D reconstruction of the interfacial sheet shown in (C), yellow box. Scanning depth 5 $\mathrm{\mu m}$. Right, sheet schematic. (F) Left, bending interfacial sheet. Centre, interfacial cup. Right, cup schematic. All scale bars, $5~\mathrm{\mu m}$.}
\end{figure*}

Although theoretical modeling, coarse-grained simulations, and reconstitution experiments are powerful tools for understanding the physical drivers of wetting phenomena in biological systems, most setups to date consider only equilibrium conditions [e.g., refs. \cite{agudo-canalejo_wetting_2021,sabet_compartmentalizing_2022,zhao_membrane_2024}]. However, the intracellular environment is intrinsically out of equilibrium and membranes continuously remodel. Recent computer simulations underpinned that the morphological changes of vesicles without interfaces are governed by non-equilibrium effects, including energy barriers and metastable states \cite{bahrami_formation_2017}. Thus, how condensate wetting and non-equilibrium effects influence the formation and diversity of membrane morphologies (tubes, sheets, and cups) remains an open question.

Here, we address this knowledge gap by combining \textit{in vivo} and \textit{in vitro} experiments with computer simulations of membranes in contact with condensates. We refer to these membrane-condensate constructs as "interfacial membranes". We report for the first time the formation of interfacial membranes in \textit{Arabidopsis thaliana} vacuoles that manifest in sheet and cup morphologies. Our findings from reconstitution experiments and simulations reveal that wetting drives the transformation of interfacial tubes into sheets and cups by controlling the energy barriers of metastable membrane shapes through forces derived from interfacial tension. We find that due to energy barriers, condensate-mediated shape transformations are hysteretic, \textit{i.e.}, depend on the history of shape transitions. Together, these results demonstrate how condensates and non-equilibrium effects dramatically alter the membrane shape-behavior, providing a clear example of the role of condensate-membrane contacts in cellular membrane remodeling. 

\section*{Results and Discussion}
\subsection*{{\textbf{\textit{In planta}}} observation of interfacial vacuole membranes}
A hallmark of seed maturation in plant embryo vacuoles is the phase-separation of storage proteins to form liquid-like condensates \cite{feeney_protein_2018}. We have previously reported that these condensates wet and deform adjacent tonoplasts (vacuolar membrane) and can also drive the formation of tonoplast-derived membrane tubes at the condensate-tonoplast interface \cite{kusumaatmaja_intracellular_2021}. In particular, we find non-tubular tonoplast-derived membranes during live-cell imaging of transgenic TPK-GFP \textit{A. thaliana} cotyledons (embryonic leaves). To identify the physiological conditions under which these interfacial membrane structures formed, we dissected cotyledons from plant seeds at different stages of maturation before the onset of seed dormancy (see Methods; Fig.~\ref{fig:figure_1}A). Consistent with previous reports \cite{feeney_protein_2018}, we observed the formation of vacuolar condensates with 3-5 $\mathrm{\mu m}$ in diameter during the mid bent cotyledon stage (Fig.~\ref{fig:figure_1}B).

During condensate-tonoplast wetting, the condensate surface partially contacts the tonoplast membranes (\textit{i.e.}, wets) and partially the lumenal solution of the vacuole, forming a liquid-liquid interface (Fig.~\ref{fig:figure_1}C). Detailed imaging of the liquid-liquid interface revealed the presence of tonoplast-derived membranes with complex, non-tubular shapes that occupy the liquid-liquid interface. We find flattened, planar membranes, as well as deformed shapes such as curved and circular morphologies (Fig.~\ref{fig:figure_1}E, F). We refer to these flattened (open) or circular (closed) tonoplast structures as interfacial "sheets" and "cups", respectively. In all observed cases, these structures remain tethered to the tonoplast membranes. Notably, sheets and cups exhibit strong fluorescence intensities, approximately twice that of the surrounding tonoplast (Fig.~\ref{fig:figure_1}D), suggesting that they consist of two closely apposed membrane bilayers that cannot be optically resolved as distinct structures due to resolution limits.

Morphologically comparable tonoplast-derived cups were previously observed by Saito et al. in germinating wild-type \textit{A. thaliana} seeds, where they were referred to as ‘bulbs’ \cite{saito_complex_2002}. These structures were later reported to occur frequently in the lytic vacuoles of various plant tissues \cite{saito_occurrence_2011}, yet their formation mechanism remains unresolved \cite{madina_vacuolar_2019}. Protein dimerization has been suggested to play a role \cite{segami_dynamics_2014}, and a different study identified a protein of unknown function that modulates bulb abundance \cite{han_regulator_2015}. 

We observed that sheets and cups localize to the liquid-liquid interface of condensates, suggesting that the interface and associated capillary forces contribute to their formation and stabilization. Physically, the interfacial tension of vacuolar condensates governs capillarity and, thus, is likely a key parameter for shaping membranes at liquid-liquid interfaces. However, since methods for measuring and manipulating interfacial tension in cells have yet to be established, the systematic study of capillarity-mediated cup formation \textit{in vivo} remains challenging.

\begin{figure*}[!t]
    \centering
    \includegraphics[width=.9\textwidth]{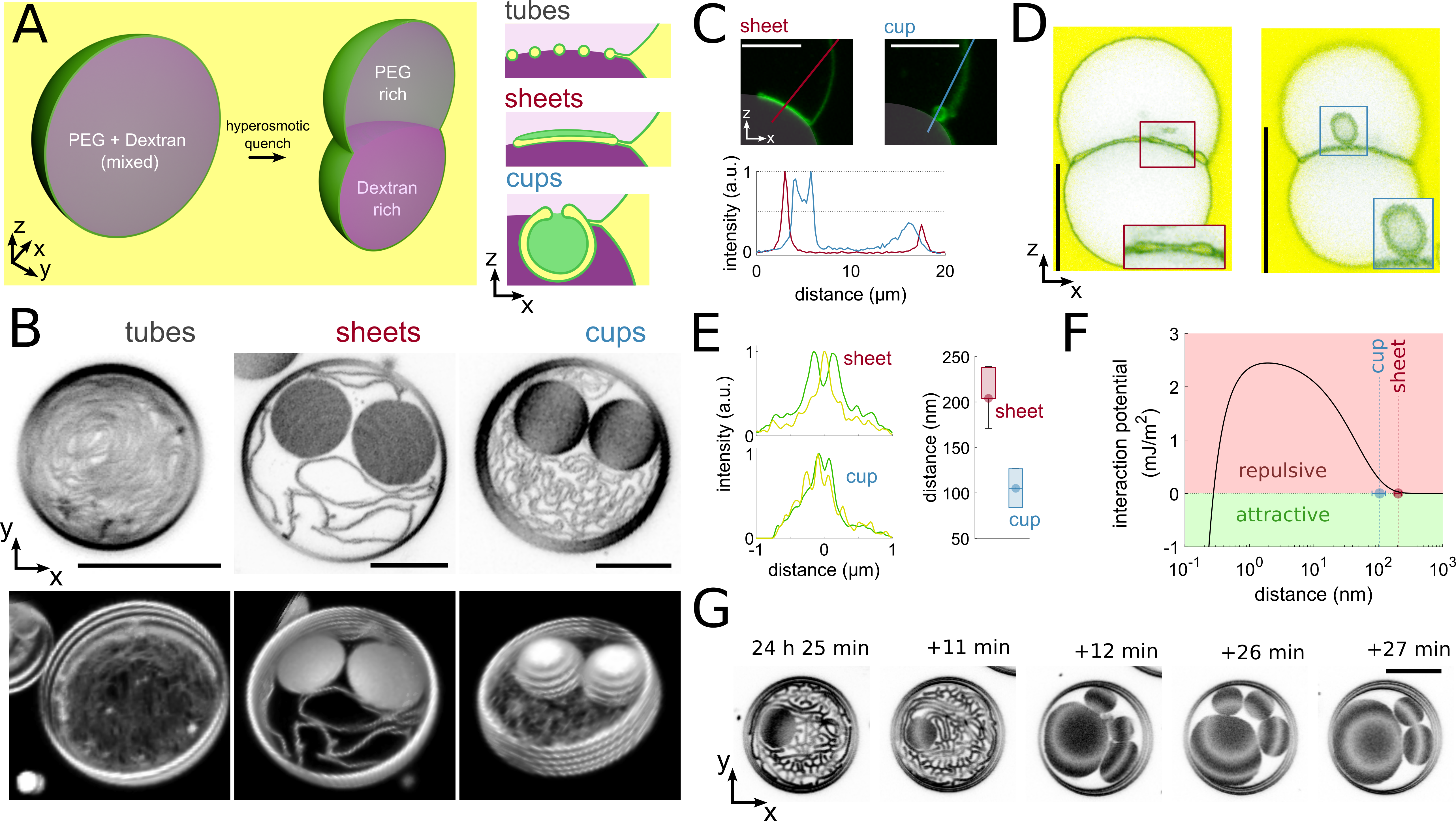}
    \caption{\label{fig:figure_2}Membrane structures at liquid-liquid interfaces inside GUV. (A) Left, schematics of GUVs (green) cut in half for illustration purposes. The GUV interior is filled with homogeneous dextran/PEG solutions (light magenta). It undergoes phase separation upon hyperosmotic quenching, forming an upper compartment of PEG-rich phase and a lower compartment of dextran-rich phase (magenta). The liquid-liquid interface is located at the equatorial neck region of the GUV. Right, interfacial membrane shapes at GUV neck. Hyperosmotic quenching generates excess membrane that accumulates at the interface in the form of tubes (top), sheets (center), and cups (bottom). (B) Confocal images of interfacial tubes, sheets, and cups. Top, maximum intensity projection. Bottom, slightly tilted 3D projection of the same interface as above.  Membrane fluorescence (DilC18) appears black in the upper projection panel and white in the lower 3D panel. (C) Sections of the interface in the xz-plane as in (A). Top left, sheet. Top right, cup. Bottom, line profiles along the arrow segments indicated above show a doubling of sheet and cup fluorescence intensity relative to the outer, unilamellar GUV membrane. (D) Sections of GUVs (green, Atto 647N-DOPE) in the xz-plane as in (A), obtained with STED microscopy. Left, sheet. Right, cup. The outer solution was stained with a water-soluble dye (yellow, Atto 488). (E) Left, averaged STED line profiles of sheets and cups as in (D), showing that the outer phase (yellow) is observed between the two membrane bilayers of sheets and cups (green). Each averaged profile was constructed with up to 10 individual line profiles, laterally cut through the adjacent bilayers, and approx. equally spaced along the structure. Right, membrane spacing of sheets ($204\pm23~\mathrm{nm}$) and cups ($105\pm25~\mathrm{nm}$), mean $\pm$ SD. (F) Interaction potential of charged membranes using Derjaguin–Landau–Verwey–Overbeek (DLVO) theory with $50~\mathrm{\mu M}$ ion concentration and $-3.2~\mathrm{mC/m^2}$ surface charge. Data points of sheets and cups from (E). Repulsive forces dominate at membrane spacing of $\leq200~\mathrm{nm}$, see SI. (G) Interfacial tubes and sheet precursors are metastable for $>24~\mathrm{h}$ before they transform within $<1~\mathrm{min}$ into several sheets that merge over time (see Movie S1). Maximum intensity projection (3 slices, $3~\mathrm{\mu m}$ scanning depth). Scale bars are $10~\mathrm{\mu m}$.}
\end{figure*}

\subsection*{{\textit{In vitro}} reconstitution of interfacial membranes} 
To study how interfacial tension, $\Sigma$, controls the shape and dynamics of interfacial membranes, we employed fluorescently labeled giant unilamellar vesicles (GUVs) as a model of plant vacuoles. To mimic intraorganelle condensate formation, we generated GUVs encapsulating a phase-separating aqueous polymer solution \cite{long_dynamic_2005} (Fig.~\ref{fig:figure_2}A). This system, which is comparable in size and mechanical properties to plant vacuoles, allows direct manipulation of $\Sigma$ by adjusting osmotic conditions and thereby the luminal polymer concentration under well-controlled settings (Fig. S1). This protein-free {\textit{in vitro}} system ensures that only physicochemical forces govern membrane wetting and shaping in the absence of confounding factors present in the complex intracellular environment. 

Using confocal microscopy, we observed interfacial membranes within the phase-separated lumen of GUVs. In these images, GUV membranes appear as ring-like structures with polarized intensity profiles (Fig.~\ref{fig:figure_2}B, C). When viewed in the $xy$-plane, the liquid-liquid interface within the GUV is initially decorated with tubular interfacial membranes similar to those found on early-stage protein condensates in vacuoles \cite{kusumaatmaja_intracellular_2021}. Within $1$ to $\geq 24$ hours, circular membrane structures emerged that coexisted with tubes. We classified these new structures as interfacial sheets and cups: while sheets exhibit homogeneous fluorescence across their planar area profile, cups display a distinct fluorescence polarization, similar to the outer GUV membrane (Fig.~\ref{fig:figure_2}B, top; Fig. S2). Three-dimensional reconstruction of z-stack images revealed quasi-planar sheet and spherical cup morphologies (Fig.~\ref{fig:figure_2}B bottom). Overall, tubes, sheets, and cups each have distinct features in equilibrium and constitute discrete and unique morphological categories.

Cross-sections in the $xz$ plane indicate that the membrane fluorescence of sheets and cups is approximately twice as strong as that of the surrounding GUV membranes (Fig.~\ref{fig:figure_2}C). As with vacuolar interfacial membranes, this indicates the presence of two bilayers in very close proximity, separated below the optical resolution limit (approx. $240~\mathrm{nm}$, see Methods). To overcome the optical resolution limit, we used stimulated emission depletion (STED) microscopy, which achieves spatial resolution down to approximately $ 50~\mathrm{nm}$ (Fig.~\ref{fig:figure_2}D–F). Importantly, averaged intensity profiles of the outer solution (yellow, membrane-impermeable dye) clearly spiked between (sheets) or on (cups) the membrane peaks (green) (Fig.~\ref{fig:figure_2}E), indicating that the outer solution fills the space between the individual bilayers. This suggests that interfacial structures form through invagination of surrounding GUV membranes. We measured that the bilayers that make up each side of a sheet or cup are $204\pm23~\mathrm{nm}$ and $105\pm25~\mathrm{nm}$ apart, respectively. We estimated that separation distances below approximately $100~\mathrm{nm}$ are associated with elevated electrostatic repulsion forces, arising from membrane charges (Fig.~\ref{fig:figure_2}F and SI). Hence, our calculations and STED measurements show that the apposing membranes are not in contact but remain separated, likely stabilized by electrostatic repulsions. These stabilizing inter-membrane interactions are consistent with the membrane-membrane separation, which lies below the optical resolution limit of conventional confocal microscopy (as observed in Figs. \ref{fig:figure_1}D and \ref{fig:figure_2}C). Taken together, microscopic imaging reveals that the interfacial structures observed in vacuoles (Fig. \ref{fig:figure_1}) and in GUVs (Fig. \ref{fig:figure_2}) show consistent structural features, suggesting that both arise from the same underlying physical principles. 

Next, we performed time-lapse imaging to monitor shape transitions in interfacial membranes (Fig. \ref{fig:figure_2}G). We initially observed a network of tubes with a small sheet-like structure at a liquid–liquid interface that had stabilized for approximately 24 hours. After remaining unchanged for about 11 minutes, the system underwent a rapid transformation within less than one minute, forming several sheets. These sheets remained largely stable, with only two small sheets undergoing subsequent coalescence. 

The coexistence of multiple morphologies (e.g., tubes, sheets, and cups; Fig. \ref{fig:figure_2}B) and the long delay before transition are indicative of a rugged energy landscape with pronounced metastability and energy barriers. Such behavior suggests that interfacial membrane shapes and their transitions are governed by non-equilibrium processes, such as kinetic trapping and slow relaxation, rather than by equilibrium thermodynamics alone.\\

\subsection*{{\textit{In silico}} model for characterizing interfacial vesicles} 

\begin{figure*}
    \centering
    \includegraphics[width=.58\textwidth]{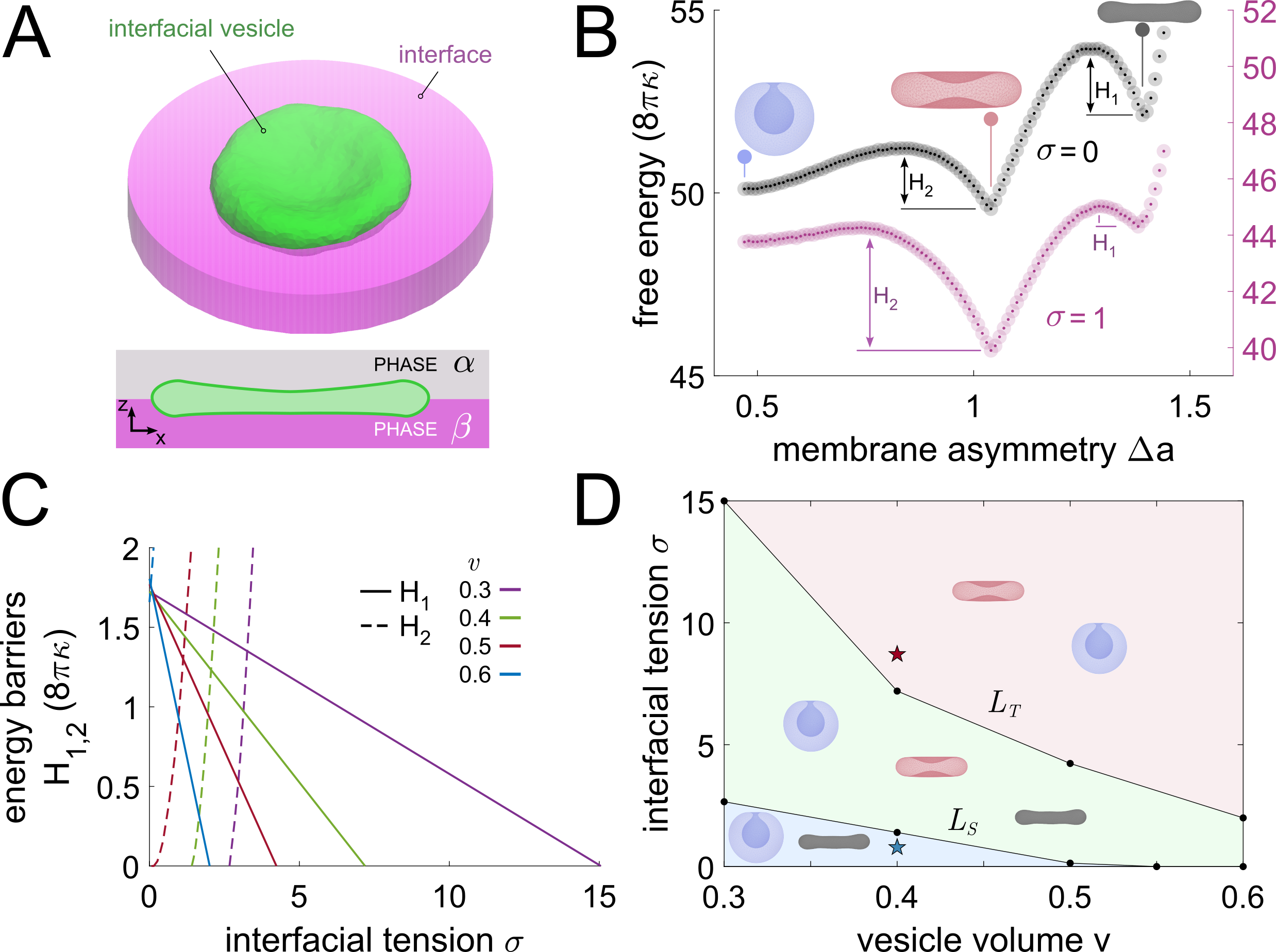}
    \caption{\label{fig:figure_3} Free energy landscape of {\textit{in silico}} vesicles. (A) Top, triangulated sheet-like vesicle (green) wetting a flat condensate surface (magenta) characterized by interfacial tension $\sigma$. Bottom, schematic of the cross-section. (B) Minimized free energy landscape of vesicles with a reduced volume $v= 0.6$ upon changing membrane asymmetry $\Delta a$. Black and left ordinate, $\sigma=0$; magenta and right ordinate, $\sigma=1$. Shape transitions between tube, sheet, and cup are associated with two energy barriers, $H_1$ and $H_2$, that change with $\sigma$. Mean $\pm$ SD for $n = 5$ independent MC minimizations. Snapshots of membranes illustrate the expected shapes at respective metastable points, irrespective of the presence or absence of an interface. (C) Energy barriers $H_1$ (solid lines) and $H_2$ (dashed lines) depend on interfacial tension $\sigma$, and their values can be obtained continuously for all $\sigma$ from our computed energy landscapes (see Fig.~S8 for specific transition values of $\sigma$). Colors correspond to $v=0.3-0.6$. (D) Morphological diagram of {\textit{in silico}} interfacial vesicles in terms of $\sigma$ and $v$ obtained based on the data in (C). $L_T$ line demarcates the area of unstable tubes (red); $L_S$ demarcates the area of unstable sheets (blue). Middle area (green) allows for all morphologies. Stars correspond to tube-to-sheet (red) and sheet-to-cup (blue) transition conditions used in Fig. \ref{fig:figure_5}A,B.}
\end{figure*}

To obtain a mechanistic understanding of how sheets and cups form at liquid-liquid interfaces, we developed a computational model to describe complex, non-equilibrium membrane morphologies at liquid-liquid interfaces. We seek a simple model that phenomenologically captures key features of the experimentally observed interfacial membrane structures, focusing only on the elastocapillary wetting physics without inter-membrane interactions. Because steady-state shapes (prolates, oblates, and stomatocytes) of the classical model of freely suspended vesicles \cite{seifert_shape_1991, seifert_configurations_1997} resemble shapes of interfacial membranes (tubes, sheets, and cups), we adopt this framework by incorporating a flat liquid-liquid interface that contacts the membrane with a wetting angle of $\theta=90^\circ$ and exerts an interfacial tension on the vesicle (Fig.~\ref{fig:figure_3}A). For our model, we employ a simple wetting approximation with an implicit liquid–liquid interface, contributing to the free energy with the area, calculated from the vesicle's projected area onto a flat interface (see Methods and SI for details). Our model is defined by two input parameters: the volume-to-area ratio of the vesicle [$v=6\sqrt \pi V/A^{3/2}$] and the reduced interfacial tension [$\sigma=\Sigma R^2/(2\kappa)$]. The latter represents the ratio between interfacial energy and elastic restoring force. At high $\sigma$, interfacial structure tends to align with the interface, whereas at low $\sigma$, vesicles adopt their minimal bending-energy shapes. With these dimensionless parameters, the model becomes independent of both the membrane bending stiffness $\kappa$ (typically $10^{-19}~\mathrm{J}$) and the vesicle area ($A=4\pi R^2$; see SI for further details on model assumptions). Our model neglects contributions from varying contact angles, membrane spontaneous curvature, and discontinuities at the three-phase contact line. Extended calculations confirm these effects are negligible compared to $v$ and $\sigma$ (see Methods and SI).

To calculate the free energy of such a system and assess the shape stability of interfacial vesicles, we developed two complementary {\textit{in silico}} approaches for triangulated vesicles with well-defined $v$ and $\sigma$ based on (i) Monte Carlo (MC) minimizations and (ii) MC simulations  \cite{kroll_conformation_1992, bahrami_formation_2017, bahrami_tubulation_2012}. While MC minimization allows us to identify the minimized free energy of vesicles in the absence of membrane thermal fluctuations \cite{kirkpatrick_optimization_1983}, MC simulations enable us to explore the free energy and the pathway of thermally driven vesicle shape transitions \cite{frenkel_understanding_2002} (see Methods). 

We used the membrane leaflet area difference, $\Delta a$, as the reaction coordinate to calculate the minimized energy of interfacial vesicles with transitional and metastable morphologies using MC minimization \cite{bahrami_formation_2017,bahrami_scaffolding_2017}. The resulting energy landscape for a vesicle with $ v = 0.6$ is shown in Fig.~\ref{fig:figure_3}B. The vesicle was modeled in both wetted (magenta, $\sigma= 1$) and condensate-free (black, $\sigma= 0$) states. The energy landscapes reveal free energy differences between the three metastable equilibrium shapes—tubes, sheets, and cups—as well as the energy barriers separating them: $H_1$ (between tubes and sheets) and $H_2$ (between sheets and cups). Introducing $\sigma$ significantly alters the energy landscape, particularly the magnitudes of $H_{1,2}$. These barriers have been shown to govern both the stability and the relative frequencies of condensate-free vesicle morphologies \cite{seifert_shape_1991, bahrami_formation_2017, bahrami_scaffolding_2017}. 

To quantitatively describe how $H_{1,2}$ depend on $v$ and $\sigma$, we systematically varied these two parameters. The simulations revealed that increasing $\sigma$ gradually reduces $H_1$ until it vanishes (Fig.~\ref{fig:figure_3}C, solid lines). This behavior indicates that the tube-to-sheet transition becomes spontaneous on interfaces with sufficiently high $\sigma$, regardless of the specific value of $v$. However, strongly deflated tubes (lower $v$) require higher $\sigma$ for $H_1$ to fully disappear. In contrast, the relationship between $\sigma$ and $H_2$, which governs the sheet-to-cup transition, exhibits the opposite trend. It is well established that condensate-free sheets with low $v$ become unstable due to negligible $H_{2}\thickapprox0$ and spontaneously transform into cups \cite{seifert_shape_1991, knorr_curvature_2012,bahrami_formation_2017}. Consistently, we observe negligible $H_{2}\thickapprox0$ for low $\sigma$ and strongly deflated sheets. However, when $v\lesssim 0.5$, significant barriers between sheets and cups ($H_2>0$) emerge once $\sigma$ becomes sufficiently large (Fig.~\ref{fig:figure_3}C dashed lines). Once such a $\sigma$-mediated $H_2$ barrier is established, $H_2$ scales proportionally with $\sigma$.

Based on the vanishing energy barriers $H_{1,2}$, we constructed a morphological diagram to visualize the stability of interfacial vesicles in the $\sigma$-$v$ parameter space (Fig.~\ref{fig:figure_3}D). The diagram delineates three distinct regions that are separated by two instability lines: $L_T$ ($H_1=0$) and $L_S$ ($H_2=0$). Below $L_S$ and beyond $L_T$ are the regimes where no sheets (blue region) and no tubes (red region) are expected, respectively. In the intermediate green region, tubes, sheets, and cups are likely to coexist. 

\subsection*{Equilibrium membrane shapes, interfacial tension, and contact angle}
\begin{figure}[!t]
    \centering
    \includegraphics[width=.3\textwidth]{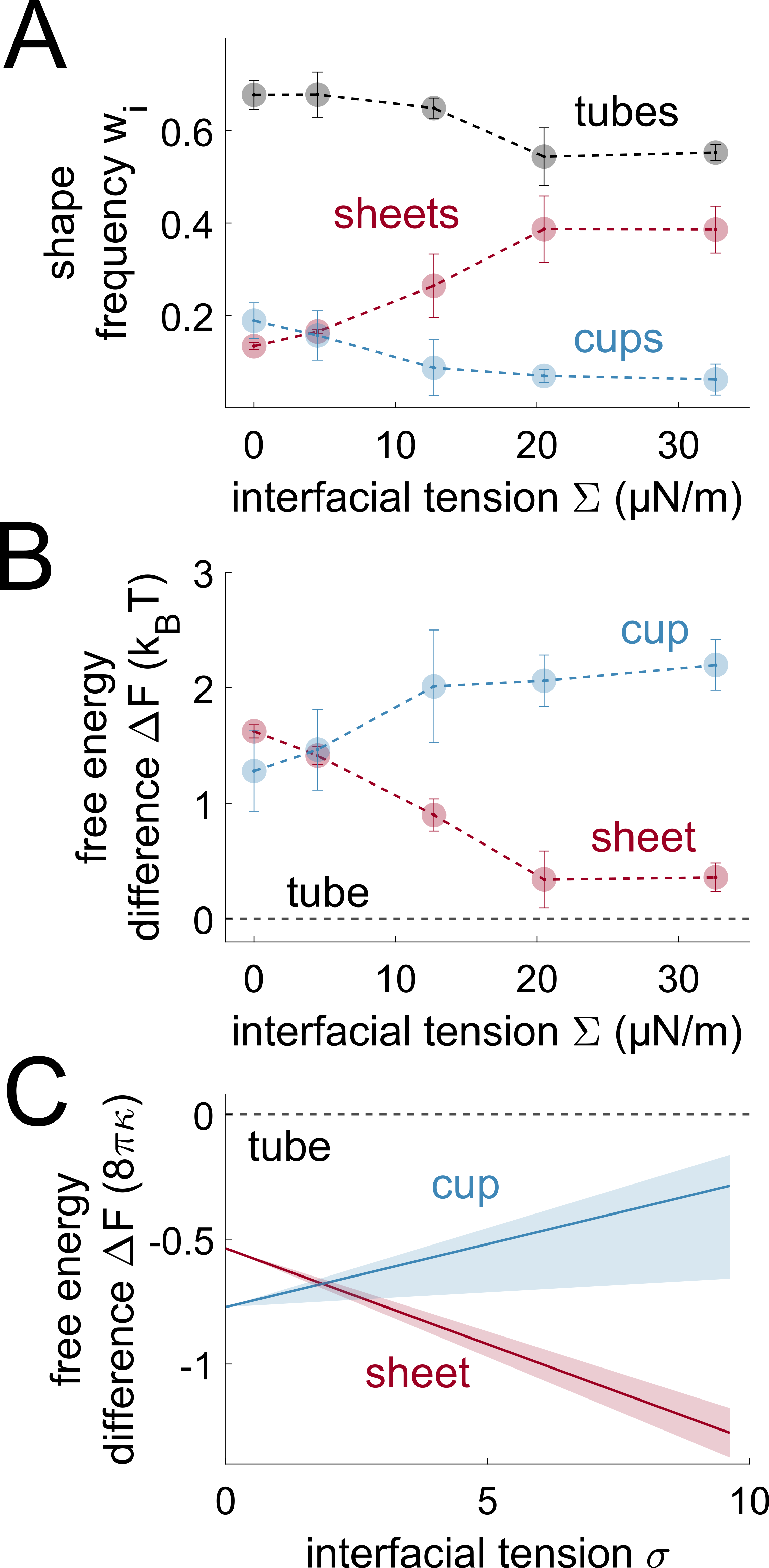}
    \caption{\label{fig:figure_4} Morphological behavior of interfacial membranes. (A) {\textit{In vitro}} shape frequencies of interfacial membranes (see Fig.~\ref{fig:figure_2}) following 24 h equilibration. (B) {\textit{In vitro}} and (C) \textit{in silico} free energy differences of cups and sheets relative to the tube state. (B) is obtained from applying \eqref{eq:energy_diff} to data presented in (A). Shaded regions correspond to free energy differences obtained from theoretical calculations with varying contact angles ($\theta=60^\circ-120^\circ$, see Fig.~S11).}
\end{figure}

We observed that shape transitions occur in a staggered, non-spontaneous manner in our reconstitution experiments (e.g, Fig. \ref{fig:figure_2}G). This behavior suggests that thermal membrane fluctuations stochastically overcome the $H_{1,2}$ energy barriers, driving shape transitions. Consequently, the height of the energy barriers likely determines how frequently shape transitions occur and sets the relative abundance of tubes, sheets, and cups at equilibrium. To understand how $\Sigma$ and $H_{1,2}$ regulate the transitions of interfacial tubes into sheets and cups \textit{in vitro}, we analyzed experimental data by quantifying the relative frequency of each morphology on interfaces with $\Sigma$ ranging from $0$ to $33~\mathrm{\mu N/m}$ after $24~\mathrm{h}$. We found that cups were more frequent than sheets at low $\Sigma$, while the opposite holds at high $\Sigma$ (Fig.~\ref{fig:figure_4}A). The crossover point, at which both morphologies appear with equal frequency, occurs at $\Sigma \thickapprox 5 ~\mathrm{\mu N/m}$. Tubes, by contrast, persisted across nearly all interfaces, independent of $\Sigma$. These observations align with the $\sigma-H_{1,2}$ relation predicted by our simulations: small $H_1$ and large $H_2$ on high-$\Sigma$ interfaces stabilize sheets. Conversely, cups dominate on low-$\Sigma$ interfaces, where $H_2$ is small. Next, we calculated the free energy differences, $\Delta \mathcal{F}_i$, from the observed shape frequencies (Fig.~\ref{fig:figure_4}B; see Methods). As tubes are the initial morphology (akin to a precursor), we consider them as a reference (ground) state. The calculated free energies were compared with {\textit{in silico}} MC minimization data (Fig.~\ref{fig:figure_4}C). We find that cups, both \textit{in vitro} and \textit{in silico}, consistently exhibit lower free energies than sheets for $\Sigma\to 0~\mathrm{\mu N/m}$. However, as $\Sigma$ increases, the free energy of cups rises and that of sheets falls. Consequently, sheets become energetically more favorable than cups ($\Delta  \mathcal{F}_S<\Delta \mathcal{F}_C$) once $\Sigma$ exceeds a critical threshold. This trend is attributed to the increasing interfacial energy penalty at higher $\Sigma$: sheets cover a larger interfacial area than tubes and cups (Fig.~S3B), thereby reducing the interfacial energy penalty and the overall free energy more effectively. Overall, we find that interfacial cups are favored at low $\Sigma$, while interfacial sheets dominate at high $\Sigma$. 

Notably, we observed differences between the free energies determined \textit{in vitro} and \textit{in silico}. Most strikingly, the tubes exhibit the lowest free energy of all shapes \textit{in vitro} but the highest free energy \textit{in silico}. One reason could be our assumption of neutral wetting ($\theta=90^\circ$) in the model, while in experiments $\theta$ may vary. However, reliable measurements of $\theta$ are optically challenging due to the small size of these membrane structures. To test the neutral wetting assumption, we developed an analytical model imposing variable $\theta$ on metastable membrane shapes (see Methods and SI). We calculated the free energy differences, $\Delta\mathcal{F}_S$ and $\Delta\mathcal{F}_C$, for $v=0.3-0.6$ and $\sigma = 0-10$, as a function of $\theta$ in the range $60^\circ$–$120^\circ$ (Figs.~\ref{fig:figure_4}C, S11B). This $\theta$ range aligns with electron microscopy images that we conducted on high-pressure frozen plant vacuoles (Fig.~S4). Our analyses showed that the morphological stability of interfacial sheets is largely robust to variations in $\theta$, consistent with our previous findings from equilibrium models \cite{agudo-canalejo_wetting_2021}. In contrast, the free energy of cup-like morphologies is more sensitive to $\theta$, particularly at higher $\sigma$. This difference arises from geometry: sheets displace nearly the same interfacial area with changes in $\theta$, whereas the displaced area of cups varies markedly. Overall, we find that $\Delta\mathcal{F}_S$ and $\Delta\mathcal{F}_C$ are generally robust to changes in $\theta$, indicating that variations in $\theta$ alone are unlikely to fully account for the observed discrepancy, within the scope of the present model.

Consequently, the discrepancy of the tube-free energy in Fig. \ref{fig:figure_4} is not primarily due to the neutral wetting assumption but rather to different limitations of our model. For example, our idealized model set-up does not fully resolve the inherent complexity of the experimental interfaces: dense packing of tubes (Fig.~\ref{fig:figure_2}C) and tube-tube contacts can reduce the flexibility of tubes on the interface \cite{liu_microrheology_2006}; tubes can become entangled or form knots, preventing further shape stabilization \cite{zhao_membrane_2024}. Unlike \textit{in silico} free energies, where $v$ is a well-defined scalar, experimentally determined free energies reflect an average over a polydisperse $v$ distribution. This averaging reflects the inherent variability of reconstitution systems, particularly in the area and volume of interfacial membranes \cite{liu_patterns_2016}. Further, our MC analyses neglect spontaneous curvature effects, as equilibrium models indicate that spontaneous curvature induces only subtle shifts in morphology diagrams \cite{agudo-canalejo_wetting_2021,zhao_membrane_2024}. Nevertheless, we observed a slight preference of membranes for the PEG-rich and storage protein-rich phases, possibly introducing membrane asymmetry \cite{liu_patterns_2016}. Such asymmetry could induce subtle shifts in the energy landscape, which may become relevant at small $H_{1,2}$. For example, previous studies of \textit{in vitro} systems proposed that spontaneous curvature can stabilize nanotubes \cite{li_membrane_2011}. We also disregard inter-membrane interactions in our model, while STED measurements and DLVO-estimations of the \textit{in vitro} system indicate membrane-membrane repulsion at separations $<200~\mathrm{nm}$. Such inter-membrane interactions can cause additional shifts in the energy landscape. Overall, several simplifying model assumptions may contribute to the substantial discrepancies in the tube free energy; resolving these inconsistencies remains an open question that will require extending the current modeling framework to incorporate these and other additional effects \textit{in silico}.

Together, these data confirm that modeling of the energetic behavior of interfacial membranes using our MC simulation strategy yields qualitatively consistent results with \textit{in vitro} experiments, thereby allowing exploration of the parameters affecting shape changes during wetting events. At equilibrium, both approaches indicate that interfacial membranes/vesicles favor the sheet morphology at high $\Sigma$, whereas the cup morphology is preferred at low $\Sigma$. To better understand the equilibrium state, we next probe \textit{in situ} transformations of interfacial membranes to clarify how $\Sigma$ governs this process. 

\subsection*{Membrane shape remodeling and symmetry breaking}

MC simulations provide a detailed picture of \textit{in situ} membrane remodeling and offer insight into shape transformation pathways at temporal and spatial scales inaccessible to microscopy-based experiments. To this end, we simulated morphological transformations starting with tubular vesicles ($v=0.4$), which were equilibrated for $10^7$ MC steps on an interface with $\sigma=0$ (equivalent to a condensate-free vesicle). Consistent with Fig.~\ref{fig:figure_3}D and previous reports \cite{bahrami_formation_2017}, tubes persisted throughout the equilibration period. Upon increasing $\sigma$, tubes become unstable and start to deform (Figs.~\ref{fig:figure_5}A and S5A). For $\sigma=8.71$ (\textit{i.e.}, for vanishing $H_1$; Fig.~\ref{fig:figure_3}C red star), we observed a spontaneous breaking of axial symmetry \textit{via} lateral tube expansion, leading to the formation of a sheet-like section at about $1.5\times 10^6$ MC steps. This sheet-like section expanded quickly and absorbed the remaining area of the tube ($2 \times 10^6$ MC steps), ultimately transforming the entire structure into a biconcave sheet comprising two apposing bilayers ($3.5 \times 10^6$ MC steps). This spontaneous non-axisymmetric transition is energetically favorable, as evidenced by a continuous reduction in free energy (Fig. S6A). The transition is driven by the capillary force (proportional to $\sigma$) acting within the interface plane, perpendicular to the tube's axis of symmetry (Fig.~\ref{fig:figure_5}C top left).

Next, we decreased $\sigma<1.5$ below $L_S$, where the capillary force becomes weaker than the membrane bending force and no sheets are predicted (Fig.~\ref{fig:figure_3}D blue star). Under these conditions, we observed a curvature inversion of the downward-facing bilayer: first, the bilayer flattens ($3.6 \times 10^6$ MC steps), followed by a transition into a convex shape ($5.4 \times 10^6$ MC steps), and eventually bending of the entire sheet into a cup (Fig.~\ref{fig:figure_5}B). Interestingly, while the tube-to-sheet transition is non-axisymmetric, the sheet-to-cup transition proceeds axisymmetrically. Unlike tubes, whose axial symmetry is broken by the capillary force, sheets and cups maintain their axial symmetry owing to the stabilizing effect of the capillary force. This distinction arises from the orientation of the symmetry axis: for tubes, it lies within the interface plane, whereas for sheets and cups, it is perpendicular to the interface (Fig.~\ref{fig:figure_5}C). Therefore, maintaining or increasing $\sigma$ stabilizes sheet-like structures by preserving or further flattening their two biconcave sides (Fig.~S7).

Notably, these results suggest that the shape of interfacial membranes depends on the history of the elastocapillary force (\textit{i.e.}, $\sigma$). Both tubes and cups exist at low interfacial tension. However, sheets are required as a precursor for cups to form. Henceforth, whether the interfacial membrane adopts a tubular or cup-shaped morphology depends on the temporal evolution of the interfacial tension: it must first increase for sheets to form and subsequently decrease for cups to emerge. Thus, the tube-sheet-cup transformation exhibits morphological hysteresis. Interestingly, we find that cups are unlikely to revert to tubes. Instead, our model predicts that variations in $\Sigma$ result in reversible transitions between cups and sheets due to the high free energy (\textit{i.e.}, low energetic favorability) of tubes at $v\leq0.6$ (Fig. S8). Reversion to a tubular morphology therefore requires either $v>0.6$ or a disassembly of the interfacial membrane and subsequent reformation, for example, {\textit{via}} osmotic changes.

\subsection*{Activated shape transitions}
As discussed earlier, shape transitions beyond the instability lines where $H_{1,2} > 0$ are activated processes driven by thermal fluctuations (Fig.~\ref{fig:figure_3}D). To gain insight into the slow relaxation of metastable membrane shapes, we estimated the transition likelihood with MC simulations. For this, we determined the number of MC steps required to achieve thermally driven shape transitions, with $\sigma$ ranging from 0 to 9. Further, we complemented these results with \textit{in vitro} quantification of transition times by measuring the relative frequencies of the membrane shapes across $\Sigma= 0- 33~\mathrm{\mu N/m}$ and $t=1-24~\mathrm{h}$, and applying an analytical first-order scheme (see Methods and SI).

\begin{figure*}[!t]
    \centering
    \includegraphics[width=.9\textwidth]{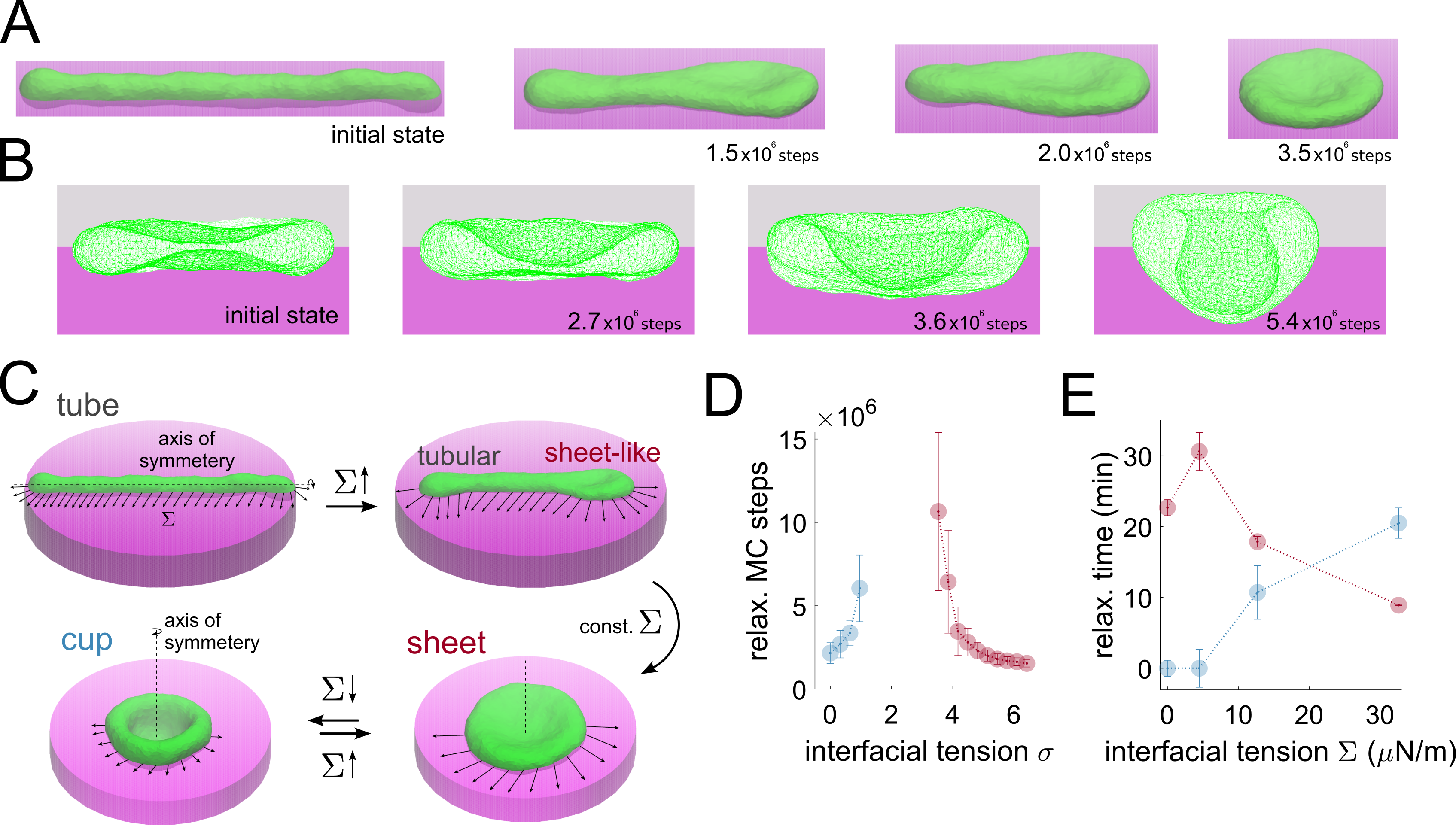}
    \caption{\label{fig:figure_5} Shape transformation pathways of interfacial vesicles (A) Isometric view of tube-to-sheet intermediate obtained by increasing $\sigma$ from 0 to 8.7. Before increasing $\sigma$, the initial tube-like, triangulated vesicle with $v=0.4$ was equilibrated for $10^7$ MC simulation steps. (B) Side view of a subsequent sheet-to-cup transition by decreasing $\sigma$ from 8.7 to 0.8. (C) Schematic of tension-induced tube-to-sheet and sheet-to-cup transitions. Increasing $\sigma$ breaks the symmetry of the initially symmetric tube and forms a sheet-like domain locally (top right). Expanding the sheet-like section adsorbs the complete tube and spontaneously transforms into a sheet {\textit{via}} an irreversible pathway at constant $\sigma$. The sheet-to-cup transition initiates upon decreasing $\sigma$ and is likely reversible, see Fig. S8. However, the whole tube-to-cup transition is hysterical: for the same $\sigma$, either tubes or cups can exist, depending on the history of the vesicle. (D) Transition times obtained from MC simulations, shown in (A,B). Mean $\pm$ SD for $n=50$ independent MC simulations. (E) {\textit{In vitro}} ensemble transition times, using the transition model (see Methods). At low $\Sigma$, the tube-to-sheet transition is slow (high $\tau_1$), while the sheet-to-cup transition is fast (low $\tau_2$). With increasing $\Sigma$, $\tau_1$ decreases (faster tube-sheet transition) while $\tau_2$ increases (slower sheet-to-cup transition).
    }
\end{figure*}

Both \textit{in vitro} and \textit{in silico} results indicate that the tube-to-sheet transition time ($\tau_1$) or equivalently the number of MC steps decreases at higher interfacial tension (Fig.~\ref{fig:figure_5}D,E red points). In contrast, high interfacial tension prolongs the sheet-to-cup transition ($\tau_2$), and requires more MC steps (Fig.~\ref{fig:figure_5}D,E blue points). \textit{In vitro}, cups formed almost exclusively and nearly instantaneously at low $\Sigma$. Both transition times converge at $\tau_1=\tau_2\approx15~\mathrm{min}$ for $\Sigma\approx 10~\mathrm{\mu N/m}$, which closely matches the sheet-cup crossover point in the equilibrium shape diagram (Fig.~\ref{fig:figure_4}A,B).

The transition times of activated processes typically scale exponentially with the corresponding energy barriers [$\tau_{1,2} \sim \exp{(-H_{1,2}})$]. The observed rapid tube-to-sheet transition (small $\tau_1$) is consistent with a reduced $H_1$, while the slow sheet-to-cup transition (high $\tau_2$) reflects the increase in $H_2$, as seen in both \textit{in vitro} and \textit{in silico} systems (Fig.~\ref{fig:figure_5}E blue points). These findings demonstrate that the characteristic trends of metastability and energy barriers are conserved across \textit{in vitro} and \textit{in silico} systems. 

Overall, our experimental data, simulations, and theoretical modeling demonstrate that condensate interfacial tension significantly influences the transition dynamics of interfacial membranes. Notably, the condensate interface also inverts the relative magnitudes of the energy barriers $H_{1,2}$ compared to those in well-studied, condensate-free membrane systems.

\section*{Conclusion}

Interactions between condensates and membranes play a crucial role in organizing the cell interior. However, their small size and highly dynamic coupling pose challenges for many experimental and computational methods. For instance, electron microscopy and molecular dynamics simulations offer high spatial resolution but limited temporal insight \cite{staehelin_nanoscale_2008, martinez-alonso_golgi_2013}, whereas live-cell imaging and mean-field models capture dynamics but sacrifice spatial detail \cite{rebane_liquid_2020, agudo-canalejo_wetting_2021}. In this study, we combine live-cell imaging of novel wetting events in plant embryo cells with reconstitution experiments and computer simulations. With this integrated approach, we bridge the limitations of individual techniques and reveal a general physical mechanism of condensate-membrane interactions that accounts for the observed phenotype in living plant embryos.

We show that the interplay between interfacial tension and the volume-to-area ratio determines the stability of interfacial membranes. This interplay can drive transitions between tube, sheet, and cup morphologies toward the most stable state. High interfacial tension promotes the rapid and robust formation of stable membrane sheets, independent of initial conditions. At moderate interfacial tensions (around $\Sigma = 5~\mathrm{\mu N/m}$), multiple morphologies can coexist, reflecting the metastability and the presence of energy barriers between shapes. Notably, our simulations reveal that cup formation exhibits hysteresis: this morphology depends not only on the instantaneous value of $\Sigma$, but also on the system’s history of shape transitions, in which tubes transform into sheets and ultimately into cups. \textit{In vitro}, we observed cups at lower and constant interfacial tensions ($\Sigma \leq 5~\mathrm{\mu N/m}$), suggesting that the $\Sigma$-dependent energy barriers can be sufficiently small for thermal membrane fluctuations to drive transitions through the tube-sheet-cup cascade. Together, these results suggest that relatively small changes in the energetic state of the condensate-membrane system can trigger irreversible transitions to the sheet-cup morphology of interfacial membranes.

In addition to our previous work showing that condensates have an important influence on the form and function of autophagosomes \cite{agudo-canalejo_wetting_2021}, condensates are also associated with membrane remodeling in other organelles. For example, tube-like intermediates \cite{staehelin_nanoscale_2008} and sheet-like structures \cite{martinez-alonso_golgi_2013, tachikawa_golgi_2017} have been observed in the stacked Golgi complex. Interestingly, the Golgi-associated protein GM130 forms membrane-bound condensates, suggesting that interfaces between this condensate and the Golgi membrane might contribute to Golgi organization \cite{rebane_liquid_2020}. The interface-driven, sheet-stabilizing mechanism described here is therefore likely applicable to these and other biological systems where condensates interact with membranes.

We observed that tonoplast-derived sheets and cups form at condensate interfaces within vacuoles of \textit{A. thaliana} embryos during early seed development. These cup-shaped structures resemble “bulbs” previously reported in seeds emerging from dormancy and during germination \cite{saito_complex_2002}. Although the mechanisms of bulb formation have remained unclear \cite{madina_vacuolar_2019}, our findings reveal that the interfacial properties of vacuolar condensates provide a novel mechanistic basis for the formation of these structures in plant embryo vacuoles.

During seed development, the cell is continuously subjected to biochemical and physiological changes that alter the intracellular environment and the properties of condensates. For example, variations in storage protein concentration within vacuoles during seed desiccation and germination likely modulate condensate composition and consequently their material properties. In plants, environmental factors such as diurnal fluctuations in temperature and humidity further influence the thermodynamic state and material properties of condensates. The varying interfacial properties give rise to a dynamic elastocapillary force balance, driving the membrane rearrangements described in this study. We propose that such shifts dictate the morphology and dynamic transitions of vacuolar interfacial membranes during both embryogenesis and germination.

As metabolically active entities, cells persist in a constant non-equilibrium state. Condensate properties and behavior vary continuously in response to metabolic, translational, stress, and regulatory states. Each of these factors affects the physicochemical environment of the cell interior and influences the intrinsic propensity of biomolecules to undergo phase separation. We speculate that temporal variations in condensate properties drive the formation of atypical morphologies of interfacial membranes, including sheets and cups. Furthermore, we provide a quantitative physical framework describing how such membrane shape transitions occur in cells. Together, our elucidation of the interplay between non-equilibrated condensate interfaces and membranes offers new insight into the dynamic remodeling of membranes within the ever-changing cellular environment.

\section*{Acknowledgments}
The work was funded by German Research Foundation grants to RLK (DFG Projects 460056461, 506366351), RLK was supported in part by the World Research Hub (WRH) Program of the International Research Frontiers Initiative, Tokyo Institute of Technology and by Exploratory Research for Advanced Technology (ERATO) Mizushima intracellular degradation project, the Japan Science and Technology Agency (JST) (JPMJER1702). AHB acknowledges support from the Max Planck Society within the framework of Max Planck Partner Group, and from the European Molecular Biology Organization, grant EMBO IG 5032. JFM is funded by a BBSRC Discovery Fellowship (BB/X010651/1). AIM was supported by JSPS KAKENHI Grant Number 25H01322.

\section*{Methods}
\subsection*{Plant embryo imaging}
The {\textit{Arabidopsis thaliana}} line used in this study harbored 35S:TPK1-GFP \cite{voelker_members_2006,maitrejean_assembly_2011}. Seedlings were kept in soil for at least 2 days at $4~^\circ\mathrm{C}$ in the dark, germinated, individualized, and grown at $22~^\circ\mathrm{C}$/65\,\% humidity day and $18~^\circ\mathrm{C}$/75\,\% humidity night with a 16\,h-light/8\,h-dark photoperiod for approximately 6 weeks. To isolate embryos, siliques from the lower part of the plants was opened, the seeds were released and kept in position using double-sided tape. Using a stereomicroscope, embryos at bent cotyledon stage were manually dissected and for vacuole staining, immediately incubated for at least 5 min with $20~\mathrm{\mu M}$ neutral red (NR, Sigma-Aldrich) solution. 

Confocal microscopy was performed using a Zeiss LSM800 Imager.Z2 microscope and C-Apochromat 40x/1.2 W korr or C-Apochromat 63x/1.2 W korr objectives or a comparable setup. 488 nm and 561 nm laser lines were used to excite GFP and NR, respectively. Emissions were collected as 505 to 540 nm for GFP and 565 to 615 nm for NR. The assay is robust to minor variations with respect to seedling germination (in soil or tissue culture and transfer to soil), plant growth conditions, imaging conditions, NR concentration, and staining procedure.

We tested two methods to prepare samples for electron microscopy. Chemical fixation of plant embryos with phase-separated vacuoles did not work and live-embryo chemical fixation confirmed that condensates dissolve within seconds. To better preserve condensates, we next used high pressure freezing. Isolated embryos were dispersed 5 wt\% soy lecithin solution, placed between carriers type B with a cavity height of 200 µm and frozen immediately using an EM HPM 100 (Leica). Freeze substitution was performed in acetone supplemented with 1 \% Osmiumtetroxide, 0.1 \% glutaraldehyde and 0.5 \% H$_2$O using the following automated program: 12 h at -90°C, gradual increase to -20°C over 18 h, 12 h at -20°C, gradual increase to +4°C over 3 h. At +4°C, the solution was exchanged for pure acetone. Staining was performed en bloc using acetone with 0.1 \% uranyl acetate and followed by standard epoxy resin infiltration, epoxy polymerization for 72 h, bloc trimming and cutting of 90 nm ultrathin sections. Sections were mounted on Si-wafers for scanning electron microscopy imaging using an Helios 5CX SEM (Thermo Fisher). The black spots visible in the SEM image originate from the supporting Si wafer.

\subsection*{Biomimetic vacuole model}
To mimic the membrane of vacuoles enclosing phase-separated lumina, we encapsulated aqueous polymers within GUV following a previously described approach \cite{long_dynamic_2005,su_kinetic_2023}. Briefly, this involves the encapsulation of a homogeneous aqueous two-polymer solution composing dextran (450-600 kDa from {\textit{Leuconostoc mesenteroides}}, Sigma-Aldrich, occasionally with a minor FITC-labeled fraction) and polyethylene glycol (PEG, 8 kDa, Sigma-Aldrich) at a ratio of 1.3 in GUVs. We prepared GUVs using the electroformation method \cite{angelova_liposome_1986}. In short, electroformation chambers were assembled by placing a Teflon spacer between two indium tin oxide (ITO) glass slides, each coated with a lipid film. After filling and sealing the chamber, we applied an alternating current (AC) of 5 V at 100 Hz for 1.5 h at 60\,°C. The lipid mixture contained 1-palmitoyl-2-oleoyl-glycero-3-phosphocholine (POPC, Avanti Polar Lipids, Birmingham AL) and 1-palmitoyl-2-oleoyl-sn-glycero-3-phospho-L-serine (POPS, Avanti Polar Lipids, Birmingham AL) at a 9:1 molar ratio, dissolved in chloroform. The membrane was fluorescently labeled with 1 mol\% membrane dye (DiIC18, Sigma-Aldrich, for confocal imaging, and Atto 647N-DOPE, ATTO-TEC, for STED imaging). For experiments with dyed outer solution, we added $10~\mathrm{\mu M}$ water-soluble dye (Atto 488, ATTO-TEC) in the outer solution. Following electroformation, the vesicles were diluted in hypertonic polymer solutions, with sucrose added to tune the osmolarity. This mild GUV dehydration shifts the internal polymer solution into a two-phase state across the binodal coexistence curve at 8.6 polymer wt\%, forming a lighter PEG-rich phase and a denser dextran-rich phase. These phases are separated by a horizontal liquid-liquid interface, characterized by a quench-specific interfacial tension. We determined interfacial tensions in bulk solutions using spinning drop tensiometry (Krüss, Hamburg; Fig.~S1). We image the GUVs with an FV1000, Olympus confocal microscope mounted with a UPLXAPO60x1.2/W, Olympus objective lens (theoretical resolution limit approx. $240~\mathrm{nm}$). Superresolution STED imaging was conducted on a TCS SP8 gSTED 3X, Leica microscope mounted with an HCPLAPOCS2100x1.4/OIL (theoretical resolution limit approx. $50~\mathrm{nm}$) with depletion at $775~\mathrm{nm}$ and excitation at $640~\mathrm{nm}$. We let the GUVs equilibrate for up to 24 hours at room temperature ($22^\circ \pm 1^\circ\mathrm{C}$), after which we observe membrane tubes, sheets, and cups at the liquid-liquid interface, resembling those seen in both the plant system and the simulation setup. We quantified frequency distributions of membrane shapes in 40-70 GUVs per interfacial tension setting and repeated these experiments 2-3 times independently. The assay is robust across different lipid mixtures, polymer batches, and imaging conditions. 
\subsection*{Computational model}

To characterize {\textit{in vivo}} and {\textit{in vitro}} interfacial membrane structures and to identify their morphological behavior in terms of $v$ and $\Sigma$, we use a triangulated elastic vesicle \cite{kroll_conformation_1992,julicher_morphology_1996,bahrami_tubulation_2012,bahrami_orientational_2013,bahrami_scaffolding_2017}. Shape transformations of freely suspended, triangulated vesicles can be understood within the theoretical framework of curvature elasticity \cite{canham_minimum_1970,helfrich_elastic_1973,seifert_shape_1991}, where vesicle shape is determined by minimizing the Helfrich bending energy \cite{helfrich_elastic_1973}. For a fixed vesicle area $A$, the resulting shape is determined by the vesicle volume $V$, commonly expressed as the dimensionless reduced volume $v=6\sqrt \pi V/A^{3/2} \leq 1$ \cite{seifert_shape_1991,bahrami_scaffolding_2017,bahrami_formation_2017}. As $v$ decreases from 1, the initial spherical vesicle ($v=1$) transitions into the minimum-energy shape, which falls among three metastable morphologies: a prolate or tubular vesicle (referred to as a tube), an oblate or discocyte vesicle (referred to as a sheet), and a stomatocyte or cup-shaped vesicle (referred to as a cup). tubes, sheets, and cups represent stable vesicle shapes at high, intermediate, and low values of $v$, respectively \cite{seifert_shape_1991,bahrami_scaffolding_2017,bahrami_formation_2017}. 
Although cups are globally stable for $v\lesssim 0.59$, an initial tubular vesicle must undergo two successive transitions—tube-to-sheet and sheet-to-cup—before forming a cup. While the tube-to-sheet transition is a discontinuous, first-order phase transition involving intermediate non-axisymmetric vesicle shapes, the sheet-to-cup transition is a continuous, second-order transition through axisymmetric vesicles \cite{seifert_shape_1991,bahrami_scaffolding_2017,bahrami_formation_2017}. 

Assuming negligible intrinsic curvature, the bending energy of the freely suspended vesicle is given by the surface integral:
\begin{equation}
\label{eq:HE}
E_b=2\kappa \int_A M^2 dA,
\end{equation}
where $M$ represents the mean curvature at any point on the vesicle surface and $\kappa$ denotes the bending stiffness of the membrane \cite{seifert_shape_1991}. 

Shape transformations of vesicles at the liquid-liquid interface are additionally influenced by the interfacial tension $\Sigma$, which acts on the vesicle along its contact line with the interface. The total free energy $E_b+\Sigma A_I$ comprises the vesicle bending energy, $E_b$, and the free energy, $\Sigma A_I$, associated with the interfacial tension of the surrounding interface with an area $A_I$. The interfacial area is defined as the difference, $A_I=A_r-A_p$, between the constant area, $A_r$, of the interfacial region surrounding the vesicle and the projected area, $A_p$, of the vesicle onto the interface (Fig.~S3A). Therefore, $A_p$, denotes the area of the interface covered by the vesicle. For a fixed $\Sigma$, the constant area, $A_r$, does not contribute to the variations in free energy and can thus be ignored. Consequently, the free energy of the vesicle and the interface is given by $\mathcal{F} = E_b-\Sigma A_p$. 

Each triangulated vesicle is composed of $N_v=2562$ vertices, $N_e=7680$ edges, and $N_t=5120$ triangles \cite{kroll_conformation_1992,julicher_morphology_1996,bahrami_tubulation_2012}. The triangle side length varies within the interval [$l$, $\sqrt{3}l$] \cite{bahrami_tubulation_2012}. The vertices of the triangulated vesicle are displaced to enable membrane remodeling and the edges are flipped to allow membrane fluidity. Vertices are randomly selected and displaced along a random direction with a displacement magnitude randomly chosen from the interval [$0$,$0.1l$]. All vesicles maintain a constant area, $A=4\pi R^2$, where the radius $R$ of the initial spherical vesicle serves as the characteristic length scale of the model. All simulations were performed in a constant area ensemble, with area fluctuations kept below $0.5\%$, achieved through applying a harmonic constraining potential. The vesicle shape can be characterized by two dimensionless parameters: the reduced volume $v$ and the rescaled interfacial tension $\sigma=\Sigma R^2/(2\kappa)$. Accordingly, the free energy of the vesicle at the interface can be written as:
\begin{equation}
\label{eq:FE}
\mathcal{F} = E_b- \frac{2\kappa}{R^2}\sigma A_p.
\end{equation}

We employed two simulation setups to explore equilibrium shapes of interfacial elastic vesicles and their transitions. In the first setup, we used MC minimization of the vesicle shapes to obtain the minimized free energies of the interfacial vesicles for a given pair of $v$ and $\Sigma$. Using these simulations, we constructed the morphological diagram and identified the energy barriers for the tube-to-sheet and sheet-to-cup transitions. In the second setup, we used MC simulations of the triangulated vesicles \cite{gompper_phase_1994,gompper_phase_1995} to study shape transitions between the equilibrium vesicle shapes and their free energies. All simulations were carried out using importance sampling from the Boltzmann distribution, implemented through the Metropolis MC algorithm \cite{frenkel_understanding_2002}.

\subsubsection*{MC minimization with simulated annealing}
MC minimizations were performed using simulated annealing MC \cite{kirkpatrick_optimization_1983}. Different vesicle shapes for each $v$ were obtained by varying $\Delta a$ as the reaction coordinate, subject to a constraint that maintained the specified value of $v$ \cite{bahrami_formation_2017}. The vesicle shape with the minimum $E_b$ at each $\Delta a$ was obtained {\textit{via}} simulated annealing, where the dimensionless temperature was gradually reduced to a small value ($k_BT=0.01$. At any fixed $\sigma$, Eq. \ref{eq:FE} was then used with the maximum value of the projected area $A_p$ of the minimized vesicle to find the minimized free energy landscape ($\mathcal{F}$) for given $v$ and $\sigma$ (Fig.~S8). Each point on the energy curve represents the mean of five independent minimizations with non-visible standard deviations.   

\subsubsection*{MC simulation}
Shape transitions of the vesicles were simulated using MC simulations of the elastic triangulated vesicle at dimensionless temperature $k_BT=1$ \cite{kroll_conformation_1992,bahrami_tubulation_2012,bahrami_orientational_2013}, where $T$ is the simulation temperature expressed in units of thermal energy. The system Hamiltonian,
\begin{equation}
\mathcal{E} = E_b- \Sigma A_p,\nonumber
\end{equation}
was sampled according to Boltzmann distribution, leading to the free energy of the interfacial vesicle as appears in Eq.~\ref{eq:FE}. Here, $A_p$ is calculated as half the sum of the projected areas of all triangles onto the $xy$ plane of the Cartesian coordinate system, with the $z$-axis perpendicular to the interface (see SI). 

\subsection*{Equilibrium shape frequencies and shape transitions}
In equilibrium, membrane shape frequencies follow the Boltzmann distribution, enabling the determination of their relative free energies. Accordingly, the shape frequencies are related to the free energies ($\mathcal{F}_i$), as

\begin{equation}
    \label{eq:energy_diff}
    \begin{split} 
        \Delta \mathcal{F}_i & = \mathcal{F}_i - \mathcal{F}_T \\
        & =   - k_BT\ln{\left(\frac{w_{i,\infty}}{w_{T,\infty}}\right)}, i=[T,S,C],
    \end{split}
\end{equation}

\noindent where $w_i$ is the relative frequency and $i=[T,S,C]$ corresponds to tubes, sheets, and cups, respectively.\\

To understand shape transitions, we construct a theoretical framework to measure the energy states and transition coefficients, \textit{in vitro}. We begin by considering a transition scheme, depicted by a morphology transition cascade from tube ($T$), to sheet ($S$), and ultimately to cup ($C$):

\begin{equation}
    \label{eq:morphology_transition_scheme}
    T \to S \to C.\\ 
\end{equation}  

\noindent We analyze the time-dependent relative frequency $w_i = n_i/(n_T+n_S+n_C)$ with $i\in\{T,S,C\}$, that we measure experimentally after 1, 7, and 24h (Fig. 10). We note that the total frequency of all morphologies sums to unity and remains conserved:

\begin{equation}
    \label{eq:conservation_constrain}
    w_T + w_S + w_C = 1.
\end{equation}

We assume first-order transitions among three membrane morphologies arranged in a linear transition cascade \eqref{eq:morphology_transition_scheme}, with their probabilities subject to \eqref{eq:conservation_constrain}. Based on these assumptions, we derive a linear, first-order evolution equation for the distribution functions \cite{sharma_thermal_2022}, as

\begin{equation}
    \label{eq:evolution_ode}
\partial_t \vec{w} = \underline{\underline{K}} \vec{w}.
\end{equation}

\noindent Here, we wrote the distribution functions compactly as $\vec{w}=[w_T,w_S,w_C]^T$. The transition matrix $\underline{\underline{K}}$ links the morphology states {\textit{via}} the linear transition path as

\begin{equation}
    \underline{\underline{K}}= \begin{bmatrix}
    -k_T & 0 & k_C\\
    k_T & -k_S & 0\\
    0 & k_S & -k_C \nonumber
\end{bmatrix}.
\end{equation}

\noindent The coefficients $k_T$, $k_S$, and $k_C$ denote the transition rate coefficients. Note, that the last two equations in \eqref{eq:evolution_ode} stem from the transition scheme \eqref{eq:morphology_transition_scheme}, while the first equation follows from the conservation constraint \eqref{eq:conservation_constrain}. The system of ordinary differential equations \eqref{eq:evolution_ode} can be solved by integration, yielding the closed-form evolution equation 
\begin{equation}
    \label{eq:distribution_evolution}
   \vec{w}(t) = C_0 \vec{v}_0 + C_1 \vec{v}_1 e^{-t/\tau_a} +C_2 \vec{v}_2 e^{-t/\tau_b}, 
\end{equation}

\noindent where the coefficient vectors $\underline{\underline{{V}}} = [\vec{v}_0, \vec{v}_1, \vec{v}_2]$ read 

\begin{equation}
    \label{eq:ode_vectors}
\underline{\underline{{V}}} = \begin{bmatrix}	
                        k_Sk_C & \tau_a^{-1} - k_S - k_C & \tau_b^{-1} - k_S - k_C\\ 
				        k_Ck_T & \tau_a^{-1} - k_C & \tau_b^{-1} - k_C\\
				        k_Tk_S & k_S & k_S\end{bmatrix}.
\end{equation}

The inverse relaxation times are

\begin{equation}
    \tau_{a,b}^{-1} = \frac{\gamma}{2}  \pm \sqrt{\left(\frac{\gamma}{2}\right)^2 - \lambda},\nonumber
\end{equation}

with 

\begin{equation}
    \begin{split}
    \gamma & = k_T + k_S + k_C,\\
    \lambda& = k_T k_S + k_S k_C  + k_C k_T.\nonumber
    \end{split}
\end{equation}

\noindent \eqref{eq:evolution_ode} describes a bimodal relaxation process. For $w_T(t)$ and $w_S(t)$,  $\tau_a$ and $\tau_b$ also inversely scale the relaxation magnitude per \eqref{eq:ode_vectors}.\\ 
We use \eqref{eq:distribution_evolution} to extract the transition rates $k_T,k_S,$ and $k_C$ from our measurements. Here, we utilize the steady-state relations $w_{T,\infty}/w_{S,\infty}=k_S/k_T$ and $w_{S,\infty}/w_{C,\infty}=k_C/k_S$ which can be obtained for $t\to\infty$ from \eqref{eq:evolution_ode}. This has the advantage of reducing the number of fitting parameters from three to one. To find the integration constants $C_0-C_2$, we use the initially measured distribution values at $\vec{w}\left(t=1~\mathrm{h}\right)$.\\
Considering the transition cascade \eqref{eq:morphology_transition_scheme}, we associate $\tau_1=\tau_a$ with the tube-to-sheet ensemble transition time. $\tau_b$ is the combined transition from tubes to cups and $\tau_1=\tau_b-\tau_a$ the sheet-to-cup ensemble transition time.\\

\subsection*{Theoretical vesicle model and the role of the contact angle}To explore the role of the contact angle between the membrane and the interface, we develop a theoretical vesicle model. We focus on the free energy differences between tubes, sheets, and cups, and demonstrate that, as initially assumed, the contact angle between the vesicle and the interface does not play a significant role in shaping interfacial vesicles. tubes, sheets, and cups can be simply represented by axisymmetric vesicles, which closely estimate their bending energies.

Tubes are assumed to be cylindrical vesicles capped by two spherical ends. Sheets are modeled as toroidal vesicles, enclosed on both sides by two flat circular planes, while cups are represented by two internal spherical vesicles. In our computational model used to analyze the simulations, we assume that the vesicle always adjusts itself at the interface to maximize the area, $A_p$, of the interface covered by the vesicles. This corresponds to vesicles with a contact angle of $\theta = \pi/2$ at the interface, where the vesicle area is equally distributed between the two liquid phases, $\alpha$ and $\beta$. 

In general, $\theta$ can assume different values within the range $0 \leq \theta \leq 2\pi$, where $\theta=0$ and $\theta=2\pi$ correspond to vesicles completely residing inside liquid phases $\alpha$ and $\beta$, respectively. For a given value of $\theta$, the total area of the interfacial vesicles is expressed as the sum:
\begin{equation}
\label{eq:areas}
A=A_{\alpha}+A_{\beta},
\end{equation}
which represents the proportions of the vesicle area wetted by the two liquid phases. The free energy of the vesicle, $\mathcal{F} = E_b + E_s$, comprises its bending energy and the interfacial energy, $E_s$, of the vesicle surfaces within the two liquids and the liquid-liquid interface, given by:
\begin{equation}
\label{eq:SurfE}
E_s = \Sigma_{\alpha} A_{\alpha}+\Sigma_{\beta} A_{\beta}+\Sigma(A_r-A_{p\beta}),
\end{equation}
where $A_{p\beta}$ denotes the area of the interface covered by the vesicle, obtained by projecting $A_{\beta}$ onto the interface. Here, $\Sigma_{\alpha}$ and $\Sigma_{\beta}$ represent the membrane tensions in the vesicle sections wetted by the two liquid phases. Assuming negligible line tension at the three-phase interfacial line, the membrane and interfacial tensions are related as follows:
\begin{equation}
\label{eq:YL}
\Sigma_{\beta}-\Sigma_{\alpha}=\Sigma\cos\theta.
\end{equation}
Using Eqs. \ref{eq:areas}, \ref{eq:SurfE}, \ref{eq:YL}, and $\Sigma = 8\pi\kappa\sigma / A$, while ignoring the constant terms ($0.5A(\Sigma_{\alpha}+\Sigma_{\beta}) + \Sigma A_r$) in $E_s$, the free energy of the extended model can be expressed as:
\begin{equation}
\label{eq:EFE}
\mathcal{F}_{ei} = E_{bi}-\frac{8\pi\kappa}{A}\sigma \bigg(A_{p\beta i}+0.5(A-2A_{\beta i})\cos\theta\bigg),
\end{equation}
where $i = [T, S, C]$. This simplifies to the original free energy in Eq.~\ref{eq:FE} when $\theta = \pi/2$, at which $A_{p\beta i}$ attains its maximum value, $A_p$. To evaluate the contribution of the contact angle, we calculate $E_{bi}$, $A_{\beta i}$, and $A_{p\beta i}$ for theoretical cylindrical tubes, toroidal sheets, and double-spherical cups, corresponding to $i = [T, S, C]$, respectively. The energies for the specific morphologies are derived in the SI.

\bibliography{pnas-ref}

\end{document}